\documentclass[reprint,amsmath,amssymb,aps,floatfix, superscriptaddress]{revtex4-2}
\usepackage{physics}
\usepackage{graphicx}
\usepackage{bm}
\usepackage[english]{babel}
\usepackage[utf8]{inputenc}
\usepackage{mathtools}
\usepackage{physics}
\usepackage{xcolor}
\usepackage{graphicx}
\usepackage[left=23mm,right=13mm,top=35mm,columnsep=15pt]{geometry} 
\usepackage{adjustbox}
\usepackage{placeins}
\usepackage[T1]{fontenc}
\usepackage{lipsum}
\usepackage{csquotes}
\usepackage{float}
\usepackage{hyperref}
\usepackage{braket}
\usepackage{esvect}
\usepackage{amsmath} 
\usepackage{xcolor}
\usepackage{comment}
\usepackage{graphicx,epsfig,float,pst-all}
\usepackage{mathrsfs}
\usepackage{inputenc}
\usepackage{textcomp}
\usepackage{mathtools}
\usepackage[normalem]{ulem}
\usepackage{bm}
\usepackage{wrapfig}
\usepackage{hyperref}
\usepackage[version=4]{mhchem}
\usepackage{siunitx}
\usepackage{placeins}
\usepackage[title]{appendix}
\usepackage{dcolumn}
\usepackage[normalem]{ulem}
\usepackage{xcolor}
\usepackage{mathrsfs}
\usepackage{inputenc}
\usepackage[slantedGreek]{mathpazo}

\usepackage[color=yellow]{todonotes}

\begin{document}
\newcommand{\rb}[1]{\textcolor{red}{#1}}
\newcommand{\rbout}[1]{\textcolor{red}{\sout{#1}}}

\preprint{APS/123-QED}

\title{\textit{Ab initio} calculations of diatomic constants and ro-vibrational parameters for the ground state of singly charged aluminium monohalides}
\author{Ankush Thakur}
\email{ankush\_t@ph.iitr.ac.in}
\altaffiliation{Contributed equally to the work}
\affiliation{Department of Physics, Indian Institute of Technology Roorkee, Roorkee-247667, India}
\author{Renu Bala}
\email{balar180@gmail.com}
\altaffiliation{Contributed equally to the work}
\affiliation{Institute of Physics, Faculty of Physics, Astronomy and Informatics, Nicolaus Copernicus University, Grudziądzka 5, 87-100 Toru\'n, Poland}
\author{H. S. Nataraj}
\affiliation{Department of Physics, Indian Institute of Technology Roorkee, Roorkee-247667, India}

\begin{abstract}
We report electronic, vibrational, and rotational spectroscopic parameters for the ground state, X$^2\Sigma^{+}$, of singly charged aluminium monohalides, employing single-reference coupled-cluster theory with single and double excitations (CCSD) together with the relativistic  basis sets. Higher order correlation effects coming from triple excitations are treated using perturbative CCSD(T) approach. Most of the molecular ions in the AlX$^+$ series, particularly barring the first two, have been studied here for the first time for their ground state electronic and vibrational structure. The vibrational parameters have been calculated by solving the vibrational Schr\"odinger equation utilizing potential energy curves and permanent dipole moment curves. Further, spontaneous and black-body radiation induced lifetimes have also been computed using relative energy separation and the transition dipole moments between the vibrational levels. The lifetimes of the lowest ro-vibrational states are found to be \(10.63\) s, \(40.39\) s, \(23.13\) s, \(31.26\) s, \(13.43\) s, and \(8.08\) s for the AlF$^+$, AlCl$^+$, AlBr$^+$, AlI$^+$, AlAt$^+$, and AlTs$^+$ ions, respectively. Furthermore, the rotational parameters such as Einstein coefficients and Franck-Condon factors for the lowest six vibrational states are also computed and reported in this work.\\

\begin{description}
\item[Keywords]
{\it ab initio} calculations, potential energy curves, spectroscopic constants, permanent dipole moment, quadrupole moment, dipole polarizability, transition dipole moment, vibrational and rotational parameters, transition rates, lifetimes.
\end{description}
\end{abstract}
\maketitle
\section{Introduction}
Cold and ultracold molecules provide numerous avenues for exploring fundamentals of quantum chemistry and physics~\cite{Carr_2009,Bohn_2017,Softley_2023}. These polar molecules have been utilized for a plethora of applications across various inter-related areas of research, including the study of controlled chemical reactions~\cite{Quemener_2012,Kosicki_2017,Ladjimi_2023}, anisotropic- and long-range dipole-dipole interactions~\cite{Carr_2009}, precision measurements of fundamental physical constants such as the fine structure constant~\cite{Flambaum_2007} and the proton-to-electron mass ratio~\cite{Kajita_2012,MKajita_2012}, besides fundamental symmetry tests~\cite{Kozlov_1995,Verma_2020} which play a crucial role in probing physics beyond the Standard Model of particle physics. Ultracold molecules are also known for being promising candidates for quantum computation~\cite{Mille_2002,Yelin_2006,KK-Ni_2018} and quantum simulation~\cite{Blackmore_2019}. These molecules can be produced through various methods, such as magnetic trapping with buffer-gas cooling~\cite{Hoang_2020}, sympathetic cooling~\cite{Molhave_2000}, and the formation from laser-cooled atoms via photoassociation~\cite{Andrew_2004} or Feshbach resonances~\cite{Kohler_2006}. The production of ultracold polar molecules and control of their internal quantum states using an external electric field enables measurements with remarkable precision ~\cite{Quemener_2012,Guo_2018}. Such a control can further be utilized in the investigation of new quantum states of matter and for simulating complex condensed-matter phenomena~\cite{Langen_2024}.\\

On the other hand, highly precise quantum many-body calculations can yield important electronic and vibrational parameters of ultracold molecules, such as bond dissociation energies~\cite{Fedorov_2017}, electric dipole moments~\cite{Geetha_2011}, dipole polarizabilities~\cite{Bala_2019}, and ro-vibrational energy levels~\cite{Khemiri_2013}, among others. These parameters, in turn, provide crucial insights into atom-molecule and ion-atom interactions~\cite{Ulmanis_2012,Klein2017}. Guided by the specific criteria of laser cooling~\cite{DiRosa_2004}, extensive theoretical works, based on many-body methods, have tried to unearth several promising candidates for ultracold studies among the diatomic molecules~\cite{Gao_2015,Kang_2015,Chuanliang_2018,Zeid_2022,Mostafa_2022,Fan_2024}. In particular, such an examination of the lower members of the singly charged aluminium monohalide family (AlX$^+$), \emph{viz}., AlF$^+$ and AlCl$^+$ ions, based on \textit{ab initio} calculations, has assessed their feasibility for laser cooling~\cite{Kang_2017}. Further, Fang \emph{et al.}~\cite{Fang_2024} have studied AlCl$^+$ for its thermodynamic stability across diverse interstellar environments, highlighting the astrophysical relevance of these molecular ions. Thus, it is imperative that the basic properties of such aluminium monohalides be known as precisely as possible.\\

A literature survey reveals that the experimental data exists for equilibrium internuclear distance ($R_e$), harmonic constant ($\omega_e$), and dissociation energy ($D_e$) for the AlF$^+$ ion only~\cite{Dyke_1984}. On the theoretical front, however, a few studies of singly charged aluminium monohalide ions have already been reported. For example, in Ref.~\cite{Kang_2017}, the authors have investigated the structural properties of the four low-lying electronic states of AlF$^+$ and AlCl$^+$ ions, including the potential energy curves (PECs), spectroscopic constants ($R_e$, $D_e$, $\omega_e$, anharmonic constant $\omega_e x_e$, and rotational constant $B_e$), and transition dipole moments (TDMs) between the ground and excited electronic states, using the multi-reference configuration interaction method with Davidson correction (MRCI+Q). Klein and Rosmus~\cite{Robert_1984} have also calculated these spectroscopic parameters including also the rotational-vibrational coupling constant ($\alpha_e$) as well as the permanent electric dipole moment (PDM) for the electronic ground state of the AlF$^+$ molecule, using the pseudo-natural orbital configuration interaction (PNO-CI) method with the coupled electron pair approximation (CEPA). \textit{Ab initio} studies of the four lowest electronic states of AlF$^+$ and AlCl$^+$ ions have been reported by Glenewinkel-Meyer $\textit{et al.}$~\cite{Meyer_1991} using MRCI with single and double excitations (MRCISD). The electron-spin $g$ shifts of the X$^2\Sigma^+$ state for the AlF$^+$ and AlCl$^+$ ions have been reported by Bruna and Grein~\cite{Bruna_2001}. Brites $\textit{et al.}$~\cite{Brites_2008} have analysed the PECs, spectroscopic constants, spin–orbit couplings, and also the transition moments between the ground and excited electronic states for the AlCl, AlCl$^+$, and AlCl$^{2+}$ molecular ions using MRCI+Q method. The PDMs of singly charged aluminium monohalides have been studied by Bala $\textit{et al.}$~\cite{Bala_2023} using the Kramers-restricted configuration interaction method considering single and double excitations (KRCISD). Recently, the spectral and thermodynamic properties along with the TDMs of the ground and excited states of the AlCl$^+$ ion using the internally contracted MRCI+Q method have been examined~\cite{Fang_2024}. Quite recently, Thakur $\textit{et al.}$~\cite{Thakur_2024} reported the PDMs and components of static dipole polarizabilities for the electronic ground state of AlX$^+$ ($X = $  F, Cl, Br, I, At and Ts) molecular systems at the CCSD(T) level of theory. However, to the best of our knowledge, electronic structure calculations for the heavier aluminium monohalide molecules having a single charge, \textit{viz.}, AlBr$^+$, AlI$^+$, AlAt$^+$, and AlTs$^+$ have not yet been reported and thus, this work serves both as a complement and an extension of that reported in Ref.~\cite{Thakur_2024}.\\

Thus, in the present work, our goal is to provide (i) accurate PECs, spectroscopic constants ($R_e$, $D_e$, $\omega_e$, $\omega_e x_e$, $\omega_e y_e$, $B_e$, and $\alpha_e$) and molecular properties including the PDMs, quadrupole moments (QMs), and components of static electric dipole polarizability ($\alpha_\parallel$ and $\alpha_\perp$) for the ground electronic state of AlX$^+$ molecular ions; (ii) vibrational parameters such as wavefunctions, energy levels ($E_v$), vibrationally coupled rotational constants (\(B_v\)), TDMs, spontaneous and black-body radiation (BBR) induced transition rates, and lifetimes of these ions. To execute our objective, we have employed coupled-cluster method restricted to single, double, and perturbative triple excitations, and relativistic Dyall basis set of quadruple zeta quality.\\

This manuscript is structured as follows: the theory and details of the calculations are discussed in the next Section~\ref{section2} followed by the results computed in the current work in Section~\ref{section3}, and the summary of our work in Section~\ref{section4}.\\

\section{Theory and Calculational details}\label{section2}
The electronic energy calculations reported in this work have been carried out at Dirac-Hartree-Fock (DHF), CCSD, and CCSD(T) levels of theory using DIRAC23~\cite{DIRAC} software suite. The Dirac-Coulomb Hamiltonian and a four-component wavefunction expanded using distinct basis sets for the large and small components has been used. However, to balance both the accuracy and the computational expense, the contribution of the two-electron (SS$\vert$SS) integrals is treated in an approximate manner as suggested by Visscher~\cite{Visscher1997}. Relativistic Dyall basis set of quadruple zeta quality (dyall.v4z)~\cite{Dyall_2006,Dyall_2016} have been used uniformly across the series of calculations, in conjunction with the Gaussian charge distribution for the nuclei and $C_{2v}$ molecular point group symmetry. The details of the basis functions for those atoms forming the diatomic molecules considered in this work are shown in Table~\ref{table-I}. The origin of the coordinate system is chosen to be at the Al atom. In order to further manage the computational expense, only those molecular orbitals that lie in the energy range from $-5E_h$ to $12E_h$ are considered for the correlation calculations. The active correlation space even with these limitations is still quite large as can be seen from  Table~\ref{table-II}, wherein the number of active electrons and virtual orbitals considered for the correlation calculations are shown. The diatomic constants are obtained from the PECs using the VIBROT program available in the MOLCAS package~\cite{OPENMOLCAS}.\\

\begin{table}[ht]
    \centering
    \caption{\label{table-I}Details of the atomic basis functions.}
    \begin{tabular}{c c c}
    \hline\hline
       Atom & Basis functions \\
       \hline
       Al & 24s, 14p, 3d, 2f, 1g \\
       F  & 18s, 10p, 3d, 2f, 1g \\
       Cl & 24s, 14p, 3d, 2f, 1g \\
       Br & 30s, 21p, 13d, 2f, 1g \\
       I  & 33s, 27p, 18d, 2f, 1g \\
       At & 34s, 31p, 21d, 14f, 1g \\
       Ts & 35s, 35p, 24d, 16f, 1g \\
       \hline\hline
    \end{tabular}
    \label{tab:my_label}
\end{table}

\begin{table}[htbp]
\caption{\label{table-II}
Number of active electrons and virtual orbitals in the dyall.v4z basis sets within an energy range of -5$E_h$ to 12$E_h$.} 
\begin{ruledtabular}
\begin{tabular}{ccc}
Molecule & Active electrons & Virtual orbitals  \\
\hline
AlF$^+$& 15 & 102 \\
AlCl$^+$& 15 & 115 \\
AlBr$^+$& 25 & 120 \\
AlI$^+$& 25 & 125 \\ 
AlAt$^+$& 29 & 138 \\
AlTs$^+$& 29 & 137 \\
\end{tabular}
\end{ruledtabular}
\end{table}

Choosing the internuclear axis for the molecular systems under study to be along the z-axis, we have calculated the parallel (\( \alpha_\parallel \equiv \alpha_{zz}\)) and perpendicular components (\(\alpha_{xx}\; \&\;  \alpha_{yy}\)) of dipole polarizability. The asymptotic behavior of molecular polarizability is determined by its atomic polarizabilities as~\cite{Jensen_2002}:
\begin{equation}
    \alpha^{\parallel}(R) \approx \alpha_{Al^+} + \alpha_X + \frac{4 \, \alpha_{Al^+} \,\, \alpha_X}{R^3} + \frac{4 \, (\alpha_{Al^+} + \alpha_X)\alpha_{Al^+} \,\, \alpha_X}{R^6}.
    \label{eq:1}
\end{equation}

The traceless QM tensor, $\Theta_{\alpha\beta}$, can be defined as~\cite{Buckingham_1959},
\begin{eqnarray}
\Theta_{\alpha\beta}=\frac{1}{2}\,\sum_{i}e_i(3r_{i_\alpha}\,r_{i_\beta}\,-\,r_{i}^2\delta_{\alpha\beta}),
\end{eqnarray}
where $\alpha$ and $\beta$ represent the Cartesian components, while the summation index $i$ runs over the number of charged particles in the system. The $zz$-component of the traceless QM tensor, \emph{viz}. $\Theta_{zz}$, is related to the other diagonal components by the equation, 
\begin{eqnarray}
\Theta_{zz}\,=-(\,\Theta_{xx}\,+\,\Theta_{yy}).
\end{eqnarray}
Of course, for linear molecules, $\Theta_{xx}\,=\,\Theta_{yy}$.\\

Using the PECs and PDM curves ranging from \(1\) {\AA} to \(30\) {\AA} for the ground electronic states at the CCSD(T) level of theory, we have solved the vibrational Schr\"odinger equation using Le Roy's LEVEL program~\cite{LEVEL} to obtain vibrational parameters, such as vibrational wavefunctions, energies, vibrationally coupled rotational constants, and TDMs between the vibrational levels. Additionally, we have calculated rotational parameters, including energies, Einstein coefficients and Franck-Condon factors (FCFs). The spontaneous and BBR-induced transition rates, at room temperature ($\mathrm{T}$ = \(300\) $\mathrm{K}$), are calculated using the relative energy separation and TDM values between the vibrational levels as~\cite{Kotochigova_2005},

\begin{eqnarray}{}  
\Gamma_{v, J}^{spon}\,=\,\sum\limits_{v^{''}, J^{''}}\Gamma^{emis}(v, J\,\rightarrow\,v^{''}, J^{''}) 
 \end{eqnarray}
and
 \begin{eqnarray}{} 
 \Gamma_{v, J}^{BBR}\,&=& \,\sum\limits_{v^{''}, J^{''}}\bar{n}(\omega)\,\Gamma^{emis}(v, J\,\rightarrow\,v^{''}, J^{''})\nonumber\\
 &+&\sum\limits_{v^{'}, J^{'}}\bar{n}(\omega)\,\Gamma^{abs}(v, J\,\rightarrow\,v^{'}, J^{'}),
  \end{eqnarray}
respectively. Here the indices ($v^{''}, J^{''}$) and ($v^{'}, J^{'}$) denote the ro-vibrational levels with energies lower and higher than that of ($v, J$) level, respectively. The average number of photons $\bar{n}(\omega)$ at frequency $\omega$ is given by the relation, 
  \begin{eqnarray}{}
  \bar{n}(\omega)\,=\,\frac{1}{e^{(\hbar\omega/k_{B}T)}-1}\,,
   \end{eqnarray}
where $\hbar\omega\,=\,\arrowvert E_{v, J}-E_{\tilde{v}, \tilde{J}}\arrowvert$ is the energy difference between the two ro-vibrational levels involved, while $k_B$ is the Boltzmann constant. Note that ($\tilde v, \tilde J$) denotes the ro-vibrational level with lower energy, i.e., ($v^{''}, J^{''}$) for emission, but is represented by the higher energy level ($v^{'}, J^{'}$) for absorption.\\

The emission and absorption rates are then calculated using the equation,
 \begin{eqnarray}{}
 \Gamma^{emis/abs} [(v, J)\,\rightarrow\,(v^{''}, J^{''})\,or\,(v^{'}, J^{'})]\,=\nonumber\\
 \,\frac{8\pi}{3\epsilon_0}\frac{\omega^3}{h c^3}\, [TDM_{(v, J)\rightarrow (v^{''}, J^{''})\,or\,(v^{'}, J^{'})}]^2.
  \end{eqnarray}
  
Finally, the total lifetimes ($\tau$) of the ro-vibrational states are obtained as,
\begin{eqnarray}{}
\tau\,=\,\frac{1}{\Gamma^{total}},
\end{eqnarray}
where $\Gamma^{total}(\,=\,\Gamma^{spon}\,+\,\Gamma^{BBR}$) is the sum of spontaneous and BBR-induced transition rates.\\
 
The Franck-Condon factors, $q_{v',v"}$, are defined as the square of the overlap matrix of the two ro-vibronic states involved~\cite{LEVEL},
\begin{eqnarray}
 q_{v',v"}\,=|\langle\Psi_{v',J'}|\Psi_{v",J"}\rangle|^2.
\end{eqnarray}

\section{Results and discussion}\label{section3}
\begin{figure}[]
    \includegraphics[width=1.03\linewidth]{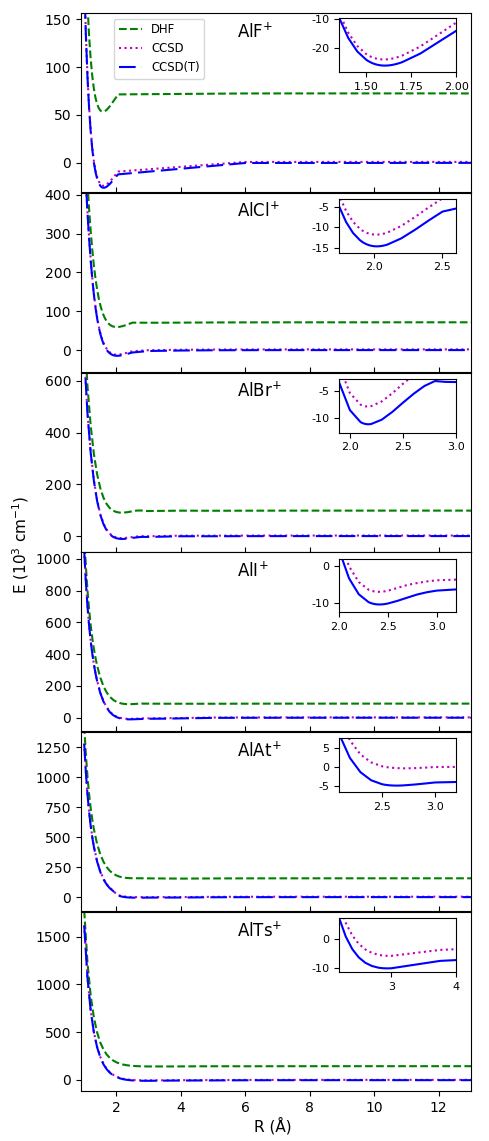}
    \caption{\label{fig:FIG1}PECs for AlX$^+$ molecular ions computed at different levels of theory using 4z basis sets, with respect to the dissociation energy at the CCSD(T) level.}
    \label{fig:1}
\end{figure}
\begin{table*}[htbp]
\begin{ruledtabular}
\begin{center}
\caption{\label{table-III}
Spectroscopic constants for the ground electronic states of the AlX$^+$ molecular ions at different levels of theory. The results computed in the present work are given in bold fonts.}
\begin{tabular}{cccccccccccccc}
 Method & $R_e$ &$D_e$  &$B_e$  & $\alpha_e$  &$\omega_e$ &  $\omega_e x_e$ &$\omega_e y_e$ & Ref.\\
 &  (\AA) & (eV) &(cm$^{-1}$) &  (cm$^{-1}$) & (cm$^{-1}$)&  (cm$^{-1}$) &(cm$^{-1}$)\\
\hline
\multicolumn{8}{c}{\textbf{AlF$^+$}}\\
\hline
 \textbf{DHF} & \textbf{1.579} & \textbf{2.356} & \textbf{0.606} & \textbf{0.0042} & \textbf{1023.814} & \textbf{4.962} & \textbf{0.106} & \textbf{This work}\\
 \textbf{CCSD} & \textbf{1.599} & \textbf{3.085} & \textbf{0.592} & \textbf{0.0044} & \textbf{976.923} & \textbf{5.310} & \textbf{0.051} & \textbf{This work}\\
 \textbf{CCSD(T)} & \textbf{1.603} & \textbf{3.212} & \textbf{0.588} & \textbf{0.0045} & \textbf{961.791} & \textbf{5.431} & \textbf{0.047} & \textbf{This work}\\
 DFT$\footnote{Using B3LYP/3z function}$ & 1.623 &-&-&-&-&-&-&~\cite{Bala_2023}\\
 MRCI+Q & 1.593 & 3.30 & 0.596 &-& 968 & 4.90 & - &~\cite{Kang_2017}\\
 RHF & 1.576 & - & 0.609 & 0.004 & 1019 & 4.2 & - &~\cite{Robert_1984}\\
 PNO-CEPA & 1.601 & - & 0.590 & 0.004 & 960 & 3.7 & - &~\cite{Robert_1984}\\
 MRCISD & 1.604 & 2.98 & 0.588 & 0.0045 & 950 & 5.57 & - &~\cite{Meyer_1991}\\
 Expt & 1.59 $\pm$ 0.01 & 3.14 $\pm$ 0.14 & - & - & 1040 $\pm$ 40 & - & -&~\cite{Dyke_1984}\\
 \hline
 \multicolumn{8}{c}{\textbf{AlCl$^+$}}\\
\hline
 \textbf{DHF} & \textbf{2.009} & \textbf{1.518} & \textbf{0.274} & \textbf{0.0014} & \textbf{643.363} & \textbf{2.417} & \textbf{0.005} & \textbf{This work}\\
 \textbf{CCSD} & \textbf{2.018} & \textbf{1.700} & \textbf{0.272} & \textbf{0.0016} & \textbf{613.488} & \textbf{2.941} & \textbf{0.044} & \textbf{This work}\\
 \textbf{CCSD(T)} & \textbf{2.023} & \textbf{1.823} & \textbf{0.271} & \textbf{0.0017} & \textbf{604.734} & \textbf{3.011} & \textbf{0.043} & \textbf{This work}\\
 DFT$\footnotemark[1]$ & 2.058 &-&-&-&-&-&-&~\cite{Bala_2023}\\
 MRCI+Q & 2.025 & 2.04 & 0.270 & - & 613.18 & 3.32 & - &~\cite{Fang_2024}\\
 MRCI+Q & 1.998 & 2.10 & 0.277 &-& 637 & 3.24 &-&~\cite{Kang_2017}\\
 MRCISD & 2.037 & 1.72 & 0.267 & 0.0015 & 592 & 2.77 &-&~\cite{Meyer_1991}\\
 MRCI+Q & 2.023 &-& 0.2703 & 0.0014 & 638.6 & 16.65 & -1.59 &~\cite{Brites_2008}\\
 \hline
 \multicolumn{8}{c}{\textbf{AlBr$^+$}}\\
\hline
 \textbf{DHF} & \textbf{2.161} & \textbf{0.976} & \textbf{0.180} & \textbf{0.0009} & \textbf{501.789} & \textbf{2.102} & \textbf{0.127} & \textbf{This work}\\
 \textbf{CCSD} & \textbf{2.168} & \textbf{1.224} & \textbf{0.178} & \textbf{0.0010} & \textbf{473.470} & \textbf{2.388} & \textbf{0.041} & \textbf{This work}\\
 \textbf{CCSD(T)} & \textbf{2.174} & \textbf{1.378} & \textbf{0.177} & \textbf{0.0010} & \textbf{466.703} & \textbf{2.336} & \textbf{0.082} & \textbf{This work}\\
 DFT$\footnotemark[1]$ & 2.223 &-&-&-&-&-&-&~\cite{Bala_2023}\\
 \hline
 \multicolumn{8}{c}{\textbf{AlI$^+$}}\\
\hline
 \textbf{DHF} & \textbf{2.390} & \textbf{0.426} & \textbf{0.133} & \textbf{0.0006} & \textbf{403.167} & \textbf{1.706} & \textbf{0.195} & \textbf{This work}\\
 \textbf{CCSD} & \textbf{2.406} & \textbf{1.124} & \textbf{0.131} & \textbf{0.0010} & \textbf{356.525} & \textbf{3.269} & \textbf{0.311} & \textbf{This work}\\
 \textbf{CCSD(T)} & \textbf{2.412} & \textbf{1.278} & \textbf{0.130} & \textbf{0.0010} & \textbf{351.999} & \textbf{3.064} & \textbf{0.258} & \textbf{This work}\\
 DFT$\footnotemark[1]$ & 2.512 &-&-&-&-&-&-&~\cite{Bala_2023}\\
 \hline
 \multicolumn{8}{c}{\textbf{AlAt$^+$}}\\
\hline
 \textbf{DHF} & \textbf{2.872} &-& \textbf{0.086} & \textbf{0.0030} & \textbf{187.125} &-&-&\textbf{This work}\\
 \textbf{CCSD} & \textbf{2.722} & \textbf{0.439} & \textbf{0.081} & \textbf{0.0366} & \textbf{173.610} & \textbf{37.620} & \textbf{14.691} & \textbf{This work}\\
 \textbf{CCSD(T)} & \textbf{2.641} & \textbf{0.607} & \textbf{0.101} & \textbf{0.0017} & \textbf{206.392} & \textbf{5.876} & \textbf{0.107} & \textbf{This work}\\
 DFT$\footnotemark[1]$ & 2.768 &-&-&-&-&-&-&~\cite{Bala_2023}\\
 \hline
 \multicolumn{8}{c}{\textbf{AlTs$^+$}}\\
\hline
 \textbf{DHF} & \textbf{3.302} & \textbf{0.403} & \textbf{0.063} & \textbf{0.0005} & \textbf{104.781} & \textbf{0.701} & \textbf{0.036} & \textbf{This work}\\
 \textbf{CCSD} & \textbf{2.987} & \textbf{0.707} & \textbf{0.076} & \textbf{0.0005} & \textbf{155.684} & \textbf{0.148} & \textbf{0.097} & \textbf{This work}\\
 \textbf{CCSD(T)} & \textbf{2.942} & \textbf{0.831} & \textbf{0.079} & \textbf{0.0006} & \textbf{167.252} & \textbf{0.558} & \textbf{0.044} & \textbf{This work}\\
 DFT$\footnotemark[1]$ & 2.928 &-&-&-&-&-&-&~\cite{Bala_2023}\\
\end{tabular}
\end{center}
\end{ruledtabular}
\end{table*}

\begin{table*}[htbp]
\caption{\label{table-IV}
Comparison of molecular energies of ground state of AlX$^+$ ions at the dissociative limit (at CCSD(T) level) with the sum of atomic energies at the same level of theory. All energy values are given in Hartrees.} 
\begin{ruledtabular}
\begin{tabular}{ccccc}
Molecule & Asymptotic Molecular State & Sum of Atomic Energies  & Molecular dissociation energy  & \% difference \\
\hline
\multicolumn{5}{c}{\vspace{-3mm}}\\
AlF$^+$ & Al$^{+}$ + F & -341.962091 & -341.962188 & 2.82 $\times$ $10^{-5}$ \\ 
AlCl$^+$ & Al$^{+}$ + Cl & -703.395715 & -703.395875 & 2.27 $\times$ $10^{-5}$ \\ 
AlBr$^+$ & Al$^{+}$ + Br & -2847.627446 & -2847.627454 & 2.84 $\times$ $10^{-7}$ \\ 
AlI$^+$ & Al$^{+}$ + I & -7358.558475 & -7358.558519 & 5.96 $\times$ $10^{-7}$ \\ 
AlAt$^+$ & Al$^{+}$ + At & -23156.264766 & -23156.264798 & 1.40 $\times$ $10^{-7}$ \\
AlTs$^+$ & Al$^{+}$ + Ts & -53768.031886 & -53768.031880 & 1.06 $\times$ $10^{-8}$ \\
\end{tabular}
\end{ruledtabular}
\end{table*}

\subsection{Spectroscopic constants}
The PECs for all the AlX$^+$ series of molecular ions computed at different levels of theory are plotted in Figure~\ref{fig:1}. The spectroscopic constants extracted from the PECs are collected and tabulated in Table~\ref{table-III}, where the results reported in the literature are also mentioned. The spectroscopic constants for all members of the AlX$^+$ series, barring the lowest two, are reported for the first time here, to the best of our knowledge. The results for the spectroscopic parameters obtained in this work for AlF$^+$ and AlCl$^+$ are also in good agreement with the available values in the literature.\\

For the AlF$^+$ molecular ion, the relative difference between our results obtained at the CCSD(T) level and those reported in Ref.~\cite{Kang_2017} using MRCI+Q theory is: $\Delta (R_e) = 0.6\%$, $\Delta (D_e) = 2.6\%$, $\Delta (B_e) = 1.3\%$, and $\Delta (\omega_e) = 0.6\%$. These differences could be attributed to the choice of computational methods and the large configurational space considered in our calculations. Our computed CCSD(T) results for $R_e$ and $D_e$ are within the error bars reported in the experimental work~\cite{Dyke_1984}. However,  $\omega_e$ shows a relative difference of about \(3.8\)\%. Our computed results for $R_e$, $\alpha_e$, $B_e$, $\omega_e$, and $\omega_ex_e$ compare very well with the results computed by Glenewinkel-Meyer $\textit{et al.}$~\cite{Meyer_1991} at the MRCISD level, with differences not exceeding \(2.5\)\%. Our computed spectroscopic constants ($R_e$, $B_e$, and $\omega_e)$ agree remarkably well with those reported by Klein and Rosmus~\cite{Robert_1984} with the differences being less than \(0.3\)\%. Nonetheless, the $\omega_ex_e$ value, as it is derived from the PECs through fitting, is highly sensitive to the number of data points included in the fitting process, and hence, shows a considerable difference. There is no data available in the literature for \(\omega_e y_e\) to make a comparison for the AlF$^+$ molecular ion.\\

We have compared the diatomic constants for the AlCl$^+$ molecular ion computed at the CCSD(T) level with the published results calculated at the MRCI+Q~\cite{Brites_2008,Kang_2017,Fang_2024} and MRCISD~\cite{Meyer_1991} levels of theory. The relative difference between the values reported in our work and those available in the literature ranges from \(0\) $-$ \(1.2\)\% for $R_e$, \(0.3\) $-$ \(2.1\)\% for $B_e$, and \(1.4\) $-$ \(5.3\)\% for $\omega_e$. The $\alpha_e$ values computed in this work are comparable to the data reported in Refs.~\cite{Meyer_1991} and \cite{Brites_2008}. The value of $D_e$ reported in our work is \(0.28\) eV smaller and \(0.1\) eV larger than those reported in Ref.~\cite{Kang_2017} and Ref.~\cite{Meyer_1991}, respectively. The anharmonic constants ($\omega_ex_e$ and $\omega_ey_e$) reported in our work differ significantly from those reported by Brites $\textit{et al.}$~\cite{Brites_2008}; however, their reported constants also differ from those obtained in the literature at a similar level of theory~\cite{Kang_2017}. \\

The PECs in Figure~\ref{fig:FIG1} show that the potentials become progressively shallower from the lighter to the heavier molecular systems, with the dissociation energy gradually decreasing as well, except for the AlTs$^+$ ion. We have also compared the $R_e$ values for all the molecular ions with the results computed using the B3LYP functional in density functional theory (DFT) by Bala $\textit{et al.}$~\cite{Bala_2023}. The maximum difference of about \(4.5\)\% found between the two works could be attributed to the difference in the methods, and also the size of the basis set employed for geometry optimization. The current work uses a much larger 4z basis set as against their 3z basis set for geometry optimization. We have observed that $B_e$ and $\omega_e$ decrease from lighter to heavier molecular ions due to an increase in bond length and reduced mass. The inclusion of correlation effects results in an increase in $R_e$ values for molecules from AlF$^+$ through AlI$^+$, however, for the heavier members such as AlAt$^+$ and AlTs$^+$, $R_e$ decreases which is likely due to strong relativistic effects in these molecular ions, as also observed for sixth and seventh row elements in Refs.~\cite{Nataraj_2010,Pyper_2020}. The PEC for AlAt$^+$ molecular ion seems to be unstable at the DHF level, especially at large internuclear distances. Therefore,  the dissociation energy and anharmonic constants for this system at this level of theory are not reported here. As we have employed a large (4z) basis set and include a large number of virtuals in our computations, the basis set superposition error in the energy calculations is expected to be negligible and should not have a significant impact on our computed results~\cite{Belpassi_2008,Miliordos_2015}.\\

To understand the dissociative nature of the molecular ions studied in this work, we have compared the molecular energies at the dissociative limit ($=30${\AA}) with the sum of atomic energies, as shown in Table~\ref{table-IV}. We have found that the relative percentage difference between them is insignificant, which confirms that AlX$^+$ molecular ions dissociate into Al$^+$ and $\text{X}$ atoms at the dissociative limit. The nature of the asymptotic molecular states for AlF$^+$ and AlCl$^+$ reported in this work agrees well with those reported in Ref.~\cite{Dyke_1984} and Refs.~\cite{Fang_2024,Kang_2017}, respectively. \\

\subsection{Molecular Properties} 
\begin{figure}[]
    \includegraphics[width=1.03\linewidth]{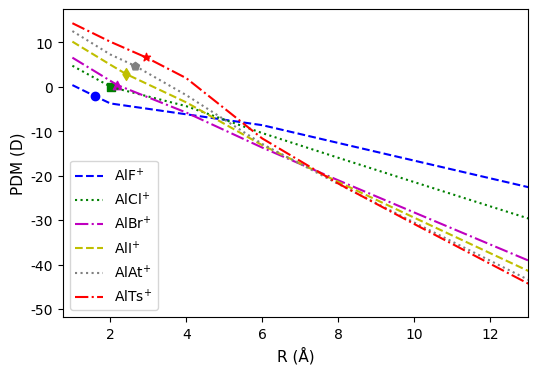}
    \caption{\label{fig:FIG2}PDMs for AlX$^+$ molecular ions computed at CCSD(T) level of theory using 4z basis sets. The markers indicate values of PDMs at equilibrium bond lengths.}
    \label{fig:enter-label}
\end{figure}

The graphs showing the behavior of PDMs of the AlX$^+$ molecular ions obtained at the CCSD(T) level, are plotted as a function of the internuclear distance in Figure~\ref{fig:FIG2}. Here, we observe that the absolute values of the PDMs beyond the equilibrium bond length increase with increasing internuclear distance for all molecular ions. This trend indicates that the barycenter of positive charge progressively shifts closer to the Al atom as the internuclear distance increases. This trivial behavior has been reported in the literature for other heteronuclear molecular ions as well~\cite{Fedorov_2017,Zrafi_2023}. We have also computed PDM values at the equilibrium bond length using the CCSD(T) level of theory, as indicated by the markers in Figure~\ref{fig:FIG2}, and found that the PDM is negative for the AlF$^+$ and AlCl$^+$ molecular ions, while positive for the rest   and increases progressively from the lighter to the heavier systems. The magnitudes of our computed PDM values, in Debye ($\mathrm{D}$), are \(2.12\) for AlF$^+$, \(0.05\) for AlCl$^+$, \(0.44\) for AlBr$^+$, \(2.90\) for AlI$^+$, \(4.77\) for AlAt$^+$, and \(6.63\) for AlTs$^+$. Our PDM values differ from those reported in Ref.~\cite{Thakur_2024}, despite both studies being conducted at the CCSD(T) level with the same dyall.v4z basis set and the same configuration space. This discrepancy arises solely from the differences in equilibrium bond lengths, as the two works have used different basis sets and different methods for geometry optimization; while Ref.~\cite{Thakur_2024} uses DFT with a 3z basis set, in this work we have extracted $R_e$s from PECs computed at the CCSD(T) level using a larger 4z basis set. Further, we have used a finer step size of \(0.001\) {\AA} around the equilibrium point. Only experimental measurements of those molecular properties can validate the accuracy of the results. However, such measurements have not been reported so far for the molecular ions examined in this work.\\ 

\begin{table}[htbp]
\caption{\label{table-V}
Electric quadrupole moments ($\Theta_{zz}$) in atomic units (au) for the ground electronic state of AlX$^+$ molecular ions computed in this work.}
\begin{ruledtabular}
\begin{tabular}{ccc}
Molecule & Method & $\Theta_{zz}$ (au)\\
\hline
AlF$^+$ & DHF & 4.005 \\ 
& CCSD &  4.049 \\ 
& CCSD(T) & 4.072 \\ 
\hline
AlCl$^+$ & DHF &  7.110 \\ 
& CCSD & 7.144 \\ 
& CCSD(T) & 7.164 \\ 
\hline
AlBr$^+$ & DHF &  9.267 \\ 
& CCSD &  9.277\\ 
& CCSD(T) &  9.297 \\
\hline
AlI$^+$ & DHF & 12.957\\ 
& CCSD & 12.924 \\
& CCSD(T) & 12.939 \\
\hline
AlAt$^+$ & DHF & 15.503\\ 
& CCSD & 15.359 \\
& CCSD(T) & 15.348 \\
\hline
AlTs$^+$ & DHF & 18.218 \\ 
& CCSD & 18.686 \\
& CCSD(T) & 18.672 \\
\end{tabular}
\end{ruledtabular}
\end{table}

\begin{table*}
\caption{\label{table-VI}
Comparison of molecular polarizability at the dissociation limit with the sum of atomic polarizabilities (at CCSD(T) level) for the ground state of AlX$^+$ ions. All the results in the table are in atomic units (au).}
\begin{ruledtabular}
\begin{tabular}{cccccc}
Molecule & $\alpha_{\text{Al}^{+}}$ + $\alpha_\text{X}$ & $\alpha_{R=30\,\text{\AA}}$ & \% difference (Columns 2 \& 3) & $\alpha$ derived from Eq.~(\ref{eq:1}) & \% difference (Columns 3 \& 5) \\
\multicolumn{6}{c}{\vspace{-3mm}}\\
\hline
\multicolumn{6}{c}{\vspace{-3mm}}\\
AlF$^+$  & 30.159 & 30.061$\footnote{At R=18.39\,\AA}$ & 0.323 & 30.249$\footnotemark[1]$ & 0.622 \\ 
AlCl$^+$ & 36.885 & 36.630 & 0.691 & 36.929 & 0.810 \\ 
AlBr$^+$  & 42.878 & 42.747 & 0.306 & 42.945 & 0.461 \\ 
AlI$^+$  & 54.038 & 53.543 & 0.916 & 54.145 & 1.112 \\ 
AlAt$^+$ & 61.980 & 61.370 &  0.984 & 62.115 & 1.199 \\
AlTs$^+$ & 79.473 & 79.080 & 0.494 & 79.672 & 0.743 \\
\end{tabular}
\end{ruledtabular}
\end{table*}

\begin{figure}[]
    \includegraphics[width=1.03\linewidth]{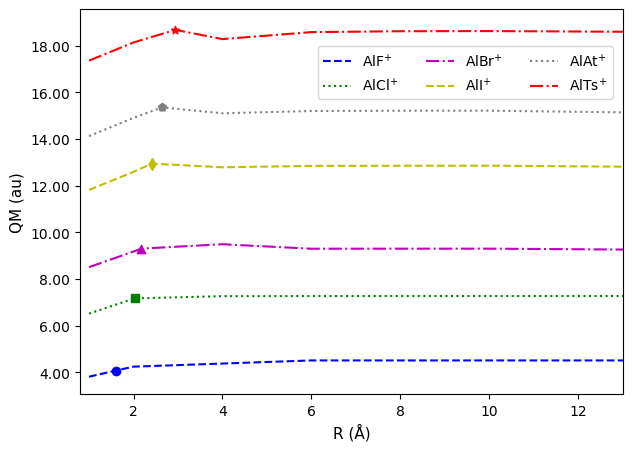}
    \caption{\label{fig:FIG3}QM curves for the AlX$^+$ molecular ions computed at CCSD(T) level of theory using 4z basis sets. The markers indicate values of QMs at equilibrium bond lengths.}
\end{figure}

\begin{figure}[]
    \includegraphics[width=1.03\linewidth]{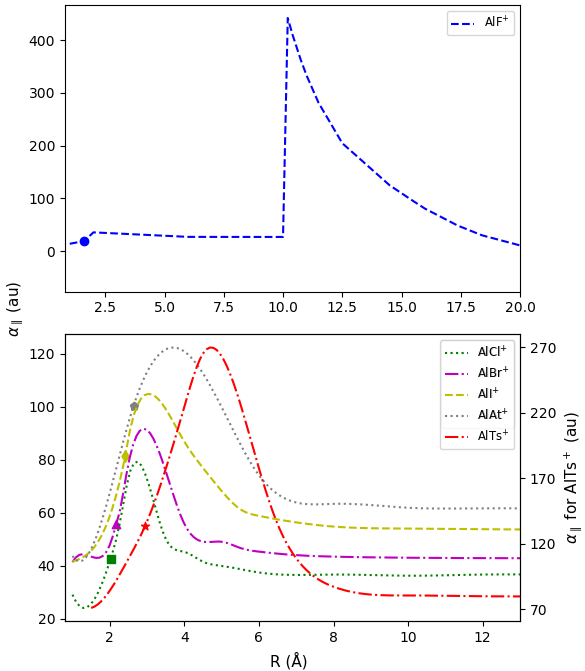}
    \caption{\label{fig:FIG4}Parallel component of static dipole polarizability as a function of internuclear distance for AlX$^+$ molecular ions computed at CCSD(T) level of theory using 4z basis sets. The markers indicate values of $\alpha_\parallel$ at equilibrium bond lengths.}
    \label{fig:enter-label}
\end{figure}

The variation of QM with internuclear distance for all molecular ions is shown in Figure~\ref{fig:FIG3}. The QM curves show more or less similar trends for all ions across the entire AlX$^+$ series. The calculated QMs are relatively small but increase progressively until the bond lengths reach their respective equilibrium distances and they attain approximately constant values at large internuclear distances. The QM values at the equilibrium bond length, evaluated at different levels of correlation, are listed in Table~\ref{table-V}. It can be seen that the magnitude of QM increases as one goes from AlF$^+$ to AlTs$^+$ molecular ions. We found that the correlation effect is most significant for the AlTs$^+$ molecular ion, with a contribution of \(\sim 2.5\)\%. To the best of our knowledge, QM results for these molecular ions are not available in the literature. Both PDMs and QMs are crucial for understanding higher-order electrostatic interactions, which exhibit complex orientation dependencies, including dipole-dipole, dipole–quadrupole, and quadrupole–quadrupole terms~\cite{Walraven_2024}.\\

The parallel component of the electric dipole polarizability increases with the internuclear distance ($R$), reaches a maximum value at $R$ > $R_e$, and then decreases to the atomic dipole polarizability limits at large $R$ values for all molecular ions, except for AlF$^+$, which exhibits a sharp peak around \(10\) {\AA}. This sharp peak could be due to avoided crossings between neighboring electronic states. These observations are clearly illustrated in Figure~\ref{fig:FIG4}. The molecular polarizability at $R = 30$ {\AA}, the sum of the atomic polarizabilities ($\alpha_{\text{Al}^{+}} + \alpha_{\text{X}}$), and the asymptotic behavior of the polarizability calculated using Eq.~(\ref{eq:1}) are tabulated in Table~\ref{table-VI}. We have found that the difference between the molecular polarizabilities at the dissociation limit and the sum of polarizabilities of the dissociating partners is not more than \(1.2\)\% among all the molecules. This further confirms the identities of the dissociative partners of AlX$^+$ molecular ions. At equilibrium bond length, our computed values for the components of dipole polarizability $(\alpha_{\parallel}, \alpha_{\perp})$ in atomic units (au) are (\(18.97\), \(21.30\)) for AlF$^+$, (\(42.60\), \(29.77\)) for AlCl$^+$, (\(55.85\), \(35.59\)) for AlBr$^+$, (\(81.40\), \(45.08\)) for AlI$^+$, (\(100.48\), \(52.61\)) for AlAt$^+$, and (\(133.17\), \(44.86\)) for AlTs$^+$ at the CCSD(T) level of theory. The increasing values in both components from AlF$^+$ to AlTs$^+$ molecular ions arise because heavier halogens have larger, more easily polarizable electron clouds, making them more responsive to electric fields. This increasing trend is consistent with the findings reported in Ref.~\cite{Thakur_2024}. \\

\begin{figure}[]
    \includegraphics[width=1.03\linewidth]{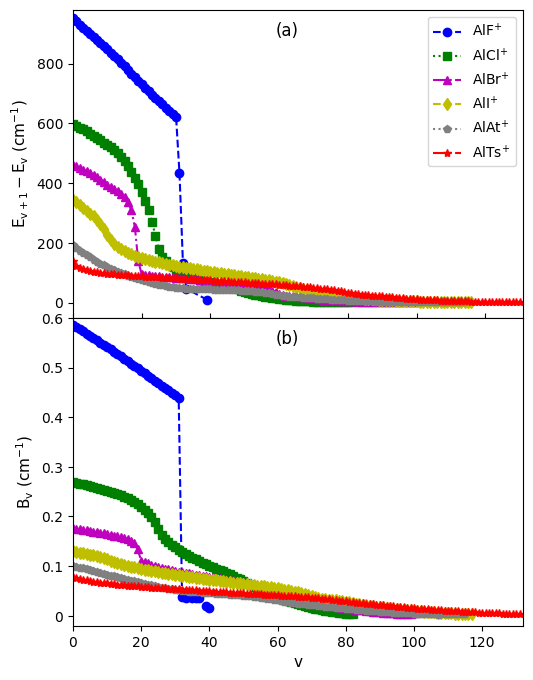}
    \caption{\label{fig:FIG5}The figure shows (a) energy spacings between adjacent vibrational levels and (b) vibrationally coupled rotational constants for AlX$^+$ molecular ions at CCSD(T) level of theory.}
\end{figure}

\begin{figure}[]
    \includegraphics[width=1.03\linewidth]{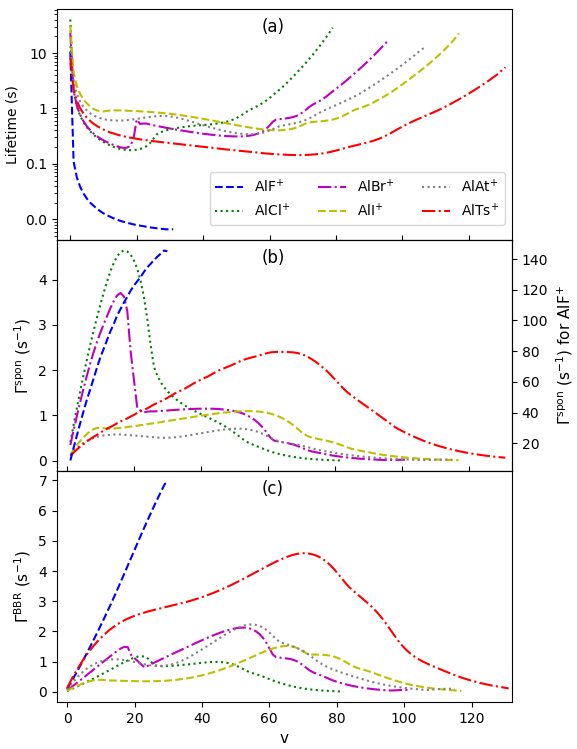}
    \caption{\label{fig:FIG6}The figure shows (a) lifetimes, (b) spontaneous transition rates, and (c) BBR-induced transition rates, at room temperature ($\mathrm{T}$ = \(300\) $\mathrm{K}$), for the vibrational levels of AlX$^+$ molecular ions.}
\end{figure}

\subsection{Vibrational parameters}
We have found \(42\), \(82\), \(101\), \(117\), \(115\), and \(131\) vibrational states for AlF$^+$, AlCl$^+$, AlBr$^+$, AlI$^+$, AlAt$^+$, and AlTs$^+$ ions, respectively. All the vibrational parameters have been obtained using the PECs and PDM curves computed at the CCSD(T) level of theory. The energy spacing between adjacent vibrational levels ($E_{v+1}-E_v$) and vibrationally coupled rotational constants ($B_v$) as a function of vibrational quantum number ($v$) are shown in Fig.~\ref{fig:FIG5}(a) and Fig.~\ref{fig:FIG5}(b), respectively. As anticipated, the relative energy separation between the vibrational levels decreases with the increase in vibrational quantum number and similar is the trend for the rotational constants. Furthermore, the vibrational energy spacings and TDMs between different vibrational levels have been used to compute the spontaneous and BBR-induced transition rates. The TDMs between different vibrational states for all the molecular ions are reported in Supplementary Table S1. From the reciprocal of the total transition rates, the lifetimes of the vibrational states are calculated. These are plotted against the vibrational quantum number, as shown in Fig.~\ref{fig:FIG6}(a). The lifetimes of the lowest ro-vibrational states, \emph{viz}. $v=0$ and $J = 0$, at room temperature are found to be \(10.63\) s, \(40.39\) s, \(23.13\) s, \(31.26\) s, \(13.43\) s, and \(8.08\) s for AlF$^+$, AlCl$^+$, AlBr$^+$, AlI$^+$, AlAt$^+$, and AlTs$^+$ ions, respectively. For all singly charged aluminium monohalide molecules, except AlF$^+$, we have observed that the lifetimes of higher vibrational states are close to those of the ground state. To analyze this further, we have presented the variation of spontaneous and BBR-induced transition rates for the vibrational states of all molecular ions as a function of the vibrational quantum number in Fig.~\ref{fig:FIG6}(b) and Fig.~\ref{fig:FIG6}(c), respectively. The spontaneous and BBR-induced transition rates for all molecular ions initially increase, reaching a maximum at a specific vibrational quantum number. Beyond this point, both transition rates begin to decline (except for the AlF$^+$ ion) because the energy spacing between highly excited vibrational states becomes smaller than that of lower energy states. Consequently, the lifetimes of the vibrational levels decrease up to a certain value of $v$, i.e., $v = 31$, $18$, $16$, $62$, $54$, and $69$ for AlF$^+$, AlCl$^+$, AlBr$^+$, AlI$^+$, AlAt$^+$, and AlTs$^+$ molecular ions, respectively, after which they begin to increase (except for AlF$^+$). However, for those vibrational levels that are close to the dissociation limit, the lifetimes approach that of the ground state. Long lifetimes for highly excited vibrational states have also been reported in the literature for alkali-alkaline-earth cations~\cite{Fedorov_2017,Bala_2019} and alkaline-earth-Francium molecular ions~\cite{Zrafi_2023}. It is important to note that the long lifetimes of the higher excited vibrational states could be beneficial for several ultracold experiments~\cite{Fedorov_2017}. To the best of our knowledge, the lifetimes of the vibrational levels of singly charged aluminium monohalide molecules are presented here for the first time. \\

Additionally, we have determined the rotational parameters such as energies, Einstein coefficients, and FCFs. The transitions involving $\Delta J = -1$, $\Delta J = 0$, and $\Delta J = +1$ are associated with the $\text{P}$, $\text{Q}$ and $\text{R}$ branches, respectively. We have performed calculations for all $\text{P}$, $\text{Q}$ and $\text{R}$ branches, considering $J_{\text{max}}$ equal to \(50\) in each vibrational level from $v = 0$ to $5$. All rotational parameters are provided in Supplementary Tables S2 to S4.\\

\section{Summary}\label{section4}
Keeping in mind the applications of ultracold diatomic molecules in various fields ranging from precision measurements of fundamental constants to quantum simulation and quantum computation, and existing gaps in the literature, we have carried out the electronic structure calculations for the ground state of singly charged aluminium monohalides using state-of-the-art many-body method. Except for the first two candidates of the AlX$^+$ series, the spectroscopic parameters for other systems have been reported for the first time in this work, to the best of our knowledge. The spectroscopic constants computed in the current work are compared with those results available in the literature. Further, we have reported the PDMs, components of static electric dipole polarizabilities, and QMs, with the latter being presented for the first time. We have found that the AlX$^{+}$ molecular ions dissociate into Al$^{+}$ and $\text{X}$ atoms at the dissociative limit, as confirmed by the excellent agreement between the molecular energies of AlX$^{+}$ ions at this limit with the sum of atomic energies, as well as the molecular polarizability at the dissociation limit with the sum of atomic polarizabilities.\\

Finally, we have obtained the vibrational wave functions from the PECs and PDM curves. Using the vibrational energy spacings and TDMs between different vibrational states, we have calculated the spontaneous and BBR-induced transition rates, and consequently, the lifetimes of the vibrational states. At room temperature, the lifetimes of the X$^2\Sigma^{+}$ state in the ro-vibrational ground state are determined to be \(10.63\) s, \(40.39\) s, \(23.13\) s, \(31.26\) s, \(13.43\) s, and \(8.08\) s for AlF$^+$, AlCl$^+$, AlBr$^+$, AlI$^+$, AlAt$^+$, and AlTs$^+$ ions, respectively. The lifetimes of vibrational states are examined as a function of the vibrational quantum number, and it is found that higher vibrational states can have lifetimes comparable to those of the ro-vibrational ground state. The rotational parameters for the $\text{P}$, $\text{Q}$ and $\text{R}$ branches have also been computed in this work, considering $J_{\text{max}} = 50$ for vibrational levels $v = 0$ to $5$.\\

We hope that the accurate spectroscopic data reported in this work for the electronic and  ro-vibrational parameters as well as for the molecular properties would be beneficial for future theoretical and experimental studies on these molecular ions.\\

\begin{acknowledgments}
We would like to thank the National Supercomputing Mission (NSM) for providing computing resources of ‘PARAM Ganga’ at the Indian Institute of Technology Roorkee, implemented by C-DAC and supported by the Ministry of Electronics and Information Technology (MeitY) and Department of Science and Technology (DST), Government of India. We would like to acknowledge Piotr \.Zuchowski, Gagandeep Singh and Roman Ciury\l{}o for the critical reading of the manuscript and for providing valuable comments. R.B. was supported by Polish National Science Centre Project No. 2021/41/B/ST2/00681. The research is also a part of the program of the National Laboratory FAMO in Toruń, Poland.
\end{acknowledgments}
\bibliography{AlX+_GS}

\end{document}